%% file: serial_link.tex
\documentclass[conference]{IEEEtran}

\usepackage[utf8]{inputenc}
\usepackage[T1]{fontenc}
\usepackage{microtype,newtxtext,newtxmath} 
\usepackage{mathtools,gensymb,url}         
\usepackage[normalem]{ulem}     
\usepackage{graphicx}
\usepackage[usenames]{color}
\usepackage{subcaption}
\usepackage{array}
\usepackage{booktabs}
\usepackage{acronym}
\usepackage[compress]{cite}
\usepackage[colorinlistoftodos]{todonotes}
\usepackage{listings}
\usepackage[switch]{lineno}
\usepackage[bookmarks={false},colorlinks,breaklinks]{hyperref} 

\hypersetup{hypertexnames=true,
  linkcolor=blue,anchorcolor=black,citecolor=blue,urlcolor=blue
}

\graphicspath{{./}{imglocal/}{./img/}}
\newcommand{\revised}[3]{#2}

\input{acronyms.tex}

\begin{document}

\title{A clock-less ultra-low power bit-serial {LVDS} link for Address-Event multi-chip systems}
%

\author{\IEEEauthorblockN{Ning Qiao\\}
\IEEEauthorblockA{Institute of Neuroinformatics\\University of Zurich and ETH Zurich\\Zurich, Switzerland\\
Email: qiaoning@ini.uzh.ch}
  \and
  \IEEEauthorblockN{Giacomo Indiveri}
\IEEEauthorblockA{Institute of Neuroinformatics\\University of Zurich and ETH Zurich\\Zurich, Switzerland\\
Email: giacomo@ini.uzh.ch}
}
\maketitle

\begin{abstract}
  We present a power efficient clock-less fully asynchronous bit-serial \ac{LVDS} link with event-driven instant wake-up and self-sleep features, optimized for high speed inter-chip communication of asynchronous address-events between neuromorphic chips. The proposed \ac{LVDS} link makes use of the \ac{LEDR} representation and a token-ring architecture to encode and transmit data, avoiding the use of conventional large \ac{CDR} modules with power-hungry DLL or PLL circuits. 
  We implemented the \ac{LVDS} circuits in a device fabricated with a standard 0.18$\mu$m CMOS process. The total silicon area used for such block is of 0.14\,mm$^{2}$. We present experimental measurement results to demonstrate that, with a bit rate of 1.5\,Gbps and an event width of 32-bit, the proposed \ac{LVDS} link can achieve transmission event rates of 35.7\,M\,Events/second with current consumption of 19.3\,mA and 3.57\,mA for receiver and transmitter blocks, respectively. Given the clock-less and instant on/off design choices made, the power consumption of the whole link depends linearly on the data transmission rate. We show that the current consumption can go down to sub-$\mu$A for low event rates (e.g., $<$1k\,Events/second), with a floor of 80\,nA for transmitter and 42\,nA for receiver, determined mainly by static off-leakage currents.
\end{abstract}

\acresetall

%
\IEEEpeerreviewmaketitle

\section{Introduction}
%
%
%
%
\IEEEPARstart{T}{he} \ac{AER} protocol has been widely used in neuromorphic computing systems to connect multiple cores and chips together~\cite{Moradi_etal17,Park_etal16,Merolla_etal14a,Furber_etal14,Benjamin_etal14,Liu_etal14}, in single-chip devices for encoding sensory signals~\cite{Liu_Delbruck10} or for implementing spike-based learning mechanisms~\cite{Qiao_etal15,Giulioni_etal12}, and in multi-chip sensory-processing systems~\cite{Neftci_etal13,Serrano-Gotarredona_etal09,Chicca_etal07a}. By exploiting the asynchronous principle, the \ac{AER} protocol is extremely efficient for event-driven neural system in terms of power consumption and low latency. Bit-parallel AER is the most commonly used implementation, due to its ease of design and configuration. This strategy however is not scalable, as the width of the parallel bus and the power required to transmit these parallel event bits scales with the size of the network. This can become a critical issue for large scale neuromorphic systems, which typically employ multiple copies of \ac{AER} buses for routing events to multiple destinations and receiving events from multiple sources~\cite{Moradi_etal17,Merolla_etal14a,Furber_etal14}: these systems are normally arranged and tiled in 2D arrays with North-South, East-West, and possibly diagonal \ac{IO} links between them. This requires a very large pin-count and can lead to significant leakage and dynamic power consumption. Rather than using the full parallel \ac{AER} protocol, some approaches have resorted to employing a ``word-serial'' protocol, which groups multiple row addresses for a given column address to reduce pin count~\cite{Benjamin_etal14,Brandli_etal14}. However, it has been argued that one of the most efficient solutions for transmitting \ac{AER} data in terms of both speed and power consumption, is to use a bit-serial \ac{LVDS} scheme~\cite{Zamarreno-Ramos_etal13a}.


Event rates in neuromorphic systems tend to be sparse, but to have high peak values~\cite{Brandli_etal14}. As the time information is typically important, low latency is an essential requirement. Traditional \ac{LVDS} schemes are designed for continuous data transmission with power consumption that depends on clock frequency, and independent of the input data rate. In some \ac{LVDS} schemes it is possible to send idle comma characters and signal a pause in the data transmission. However, these idle states may cause loss of synchronization between transmitter and receiver, and many (e.g., in the order of hundreds) clock cycles are typically required for these lock recoveries.
Therefore, as traditional \ac{LVDS} implementations are likely to cause significant latency for event transmission, they are not suitable for \ac{AER} neuromorphic systems. Previous approaches have proposed to optimize the \ac{CDR} scheme so that the phase lock of transmitter and receiver can be recovered on the fly~\cite{Zamarreno-Ramos_etal13a}, but they required additional clock generation and synchronization circuits, such as DLL or PLL circuits, for the \ac{CDR} which are very expensive in terms of power and area requirements. The event-based nature of \ac{AER} data transmission in neuromorphic systems calls for the  development of a new fully asynchronous clock-less event-based switchable bit-serial \ac{AER} \ac{LVDS} link, that does not need clock recovery circuits. In this paper we propose a new clock-less \ac{LVDS} scheme optimized for neuromorphic systems, and demonstrate its implementation in a prototype chip, fabricated using a standard 0.18\,$\mu$m CMOS process. We show that the chip designed successfully implements the following features: \\
1) Pure asynchronous design without PLL/DLL for \ac{CDR}.\\
2) Instant on ($<$0.5\,ns) wake-up for new event data and instant off ($< $0.5\,ns) self-sleep in absence of data, maintaining low latency and low power consumption. \\
3) Sub-nW (220\,nW) static power consumption and event-rate based dynamic power consumption.\\
4) Compact layout as a building block for multi-core and multi-chip neuromorphic systems. 

The paper is organized as follows: Section~\ref{sec:encod-scheme-arch} presents the data transmission scheme and link architecture; Section~\ref{sec:circ-impl} describes the circuits implementation of the proposed bit-serial \ac{LVDS} link; Section~\ref{sec:experiment} presents the measurements made with the prototype chip and describes the experimental results;  Section~\ref{sec:conclusions} shortly concludes the work.

\section{Encoding Scheme and Architecture}
\label{sec:encod-scheme-arch}

\subsection{Data encoding}


\begin{figure}
  \centering
  \includegraphics[width=0.28\textwidth]{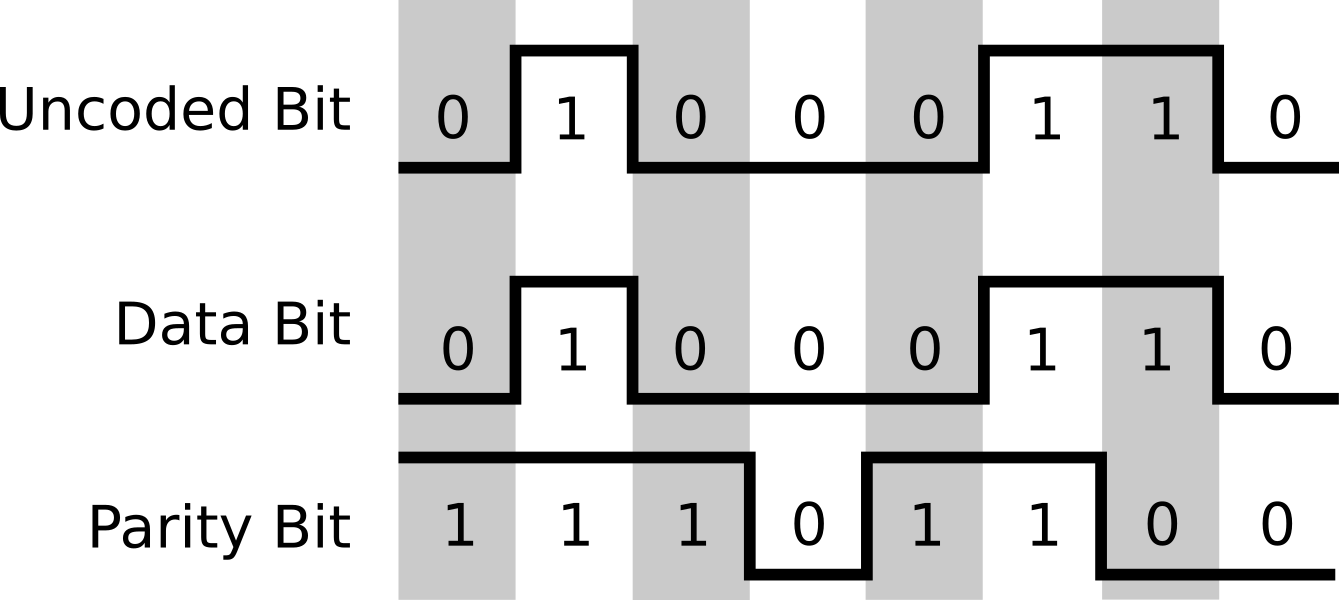}
  \caption{Encoding example of \ac{LEDR}\@. Shaded regions represent the even phase and the non-shaded regions represent the odd phase.}
  \label{fig:encoding}
\end{figure}

It is possible to implement a clock-less fully asynchronous event-driven bit-serial \ac{LVDS} link by choosing a proper data encoding scheme that eliminates the need of  traditional \ac{CDR}, which is expensive for asynchronous systems.
One scheme that is optimally suited for \ac{AER} data is the \ac{LEDR} signaling scheme~\cite{Dean_etal91}. In \ac{LEDR} signaling data bits are encoded using two rails: given a sequence of bits, a data rail is used to represent the bit value and a parity rail is used to represent the parity relative to the encoding phase and data rail. The encoding alternates between an even and an odd phase. In the even phase the parity rail takes the inverted bit value, and in the odd phase the parity rail takes the same bit value of the data rail. Formally the data rail value $D[i]$ and the parity rail value $P[i]$ are:
\[
\begin{cases}
      D[i]=B[i]; \quad P[i]=\overline{B[i]}
    & \text{for odd phase }
    \\
      D[i]=B[i]; \quad P[i]=B[i]
      & \text{for even phase }
\end{cases}
\]
where B[i] represents the encoded bit value of the sequence. Figure~\ref{fig:encoding} shows an encoding example for an 8-bit data sequence. 

The \ac{LEDR} is a \ac{DI} protocol: sequential bits can easily be distinguished by checking whether D[i]=P[i] or not. So, by encoding address event data strings using \ac{LEDR}, it is possible to build fully asynchronous bit-serial \ac{LVDS} links without using a clock generation block or a clock synchronization block for \ac{CDR}. 

According to this \ac{LEDR} encoding scheme, it is possible to implement both asynchronous encoder and decoder. On the encoder side, the data rail should always take the original sequence bit value while the parity rail should take the inverted sequence bit value for the odd phase, and the original sequence bit value for the even phase. On the decoder side, it is sufficient to check if $D[i]=P[i]$ or $P[i]=\overline{D[i]}$ to determine the bit phase, and then to read incoming bits one by one. This scheme leads to a very compact design in terms of hardware resources. Because \ac{LEDR} encoding follows a two-phase handshaking (or \ac{NRZ}) protocol, it allows a full bit rate and provides a significant bandwidth advantage comparing to alternative schemes based on \ac{PE} or \ac{DR-RZ} methods.



\subsection{LVDS with token-rings}
\label{sec:token_ring}
\begin{figure}
  \centering
  \includegraphics[width=0.45\textwidth]{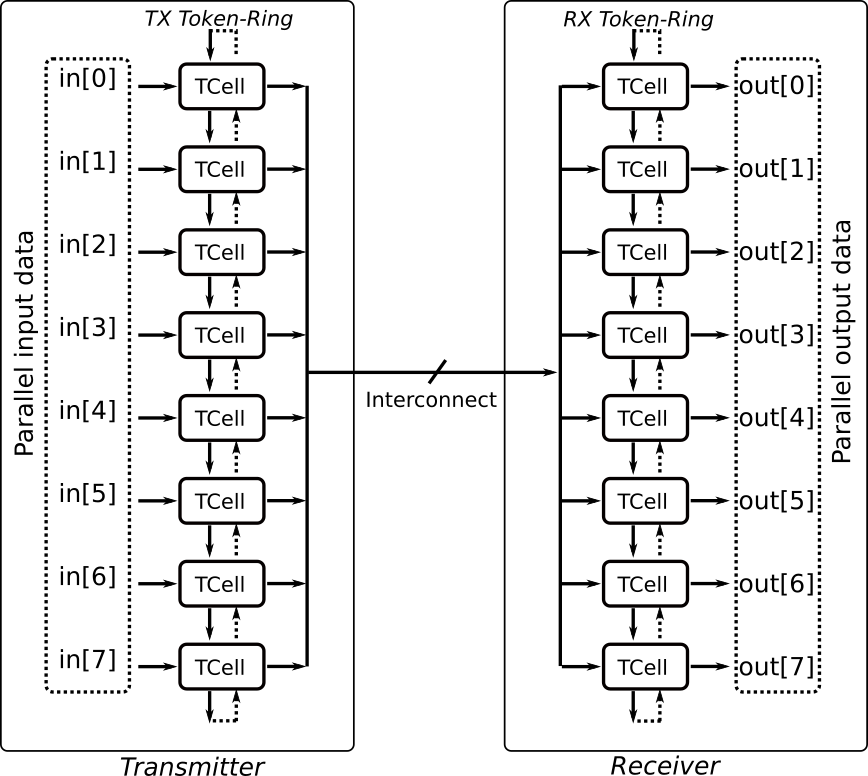}
  \caption{A typical 8-bit transceiver based on token-ring architecture. Token-cells are labeled as ``TCell''.}
  \label{fig:token_ring}
\end{figure}

\begin{figure*}[t]
  \centering
  \includegraphics[width=0.65\textwidth]{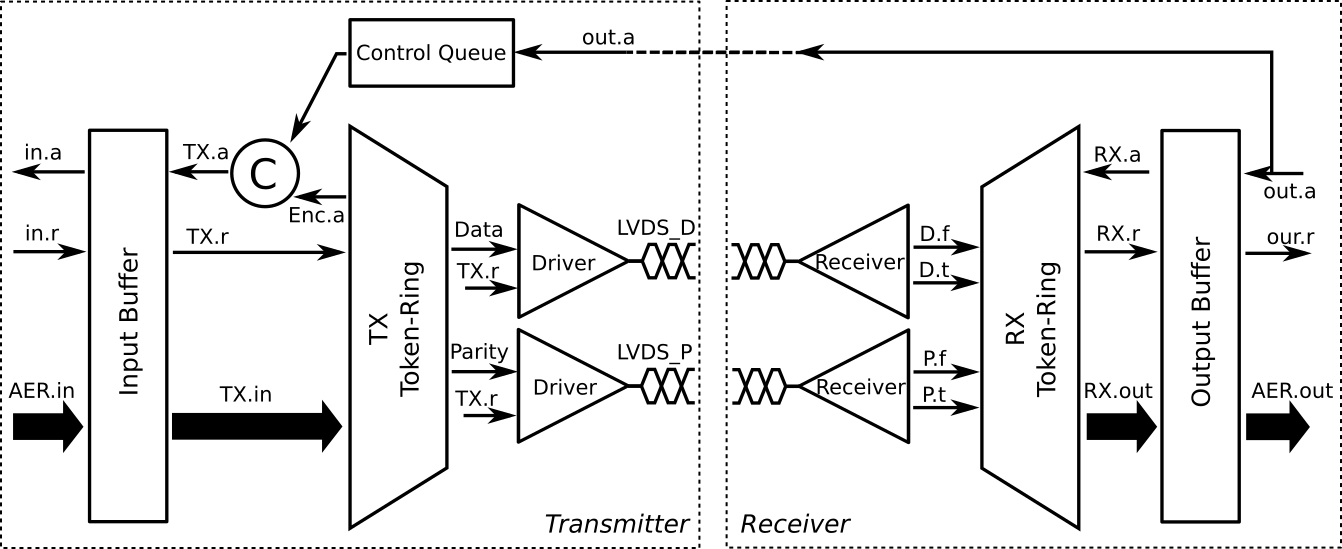}
  \caption{Architecture of the proposed bit-serial \ac{LVDS} link. }
  \label{fig:arch}
\end{figure*}

Token-ring schemes have already been proposed for asynchronous sequential data transmission~\cite{Teifel_Manohar03}. A token-ring comprises a number of mutually exclusive token-cells to transmit their data content one by one. Figure~\ref{fig:token_ring} shows a typical 8-bit transceiver based on token-ring architecture~\cite{Teifel_Manohar03}. Token-cells in the ``Transmitter block'' are activated sequentially to take one bit at a time from a parallel data bus and to write it on a shared interconnection link. Token-cells in the ``Receiver block'' take bit values sequentially from the shared bus to reconstruct the parallel data. 

A token-ring based serializer can be built following the \ac{LEDR} scheme to sequentially encode both data and corresponding parity bits from a parallel bus to a shared serial one. Accordingly, a token-ring based de-serializer can be built to de-serialize and decode the data by taking data bit-by-bit from shared data/parity wires.
The block diagram of the asynchronous bit-serial \ac{LVDS} link we propose based on these concepts is shown in Fig.~\ref{fig:arch}. It comprises the following blocks: ``Input Buffer'', ``TX Token-Ring'', ``RX Token-Ring'', ``\ac{LVDS} Drivers'', ``\ac{LVDS} Receivers'', ``Output Buffer'' and ``Control Queue''.

The ``TX Token-Ring'' block is implemented to serialize and encode event parallel bits into data and parity rails following the \ac{LEDR} scheme. The ``RX Token-Ring'' is implemented to de-serialize and reconstruct parallel bits from data and parity rails. \revised{}{The ``\ac{LVDS} Drivers'' convert data and parity rails into low-voltage differential signals for low-power consumption and high-speed inter-chip data transmission. Similarly, the ``\ac{LVDS} Receivers'' convert \ac{LVDS} signals back to normal digital signals. In order to minimize power consumption and make it depend only on the event-rate, we propose a novel instant on/off scheme for \ac{LVDS} Drivers and Receivers, described in the following section.}{} Finally, the data transmission is done in a ``burst mode'', such that the acknowledge signal is returned once per address event word, rather than bit-by-bit. Address event input and output buffers are included to pipeline the transmission cycle and increase data depth on both sides. A small ``Control Queue'' block with the same depth of the output buffer is employed to pre-store multiple acknowledges, so that the transmitter can keep on sending events without waiting for their corresponding acknowledge signals to arrive, in order to minimize latency. 

\subsection{Instant On/Off driver and receiver}

\begin{figure}
  \centering
  \includegraphics[width=0.4\textwidth]{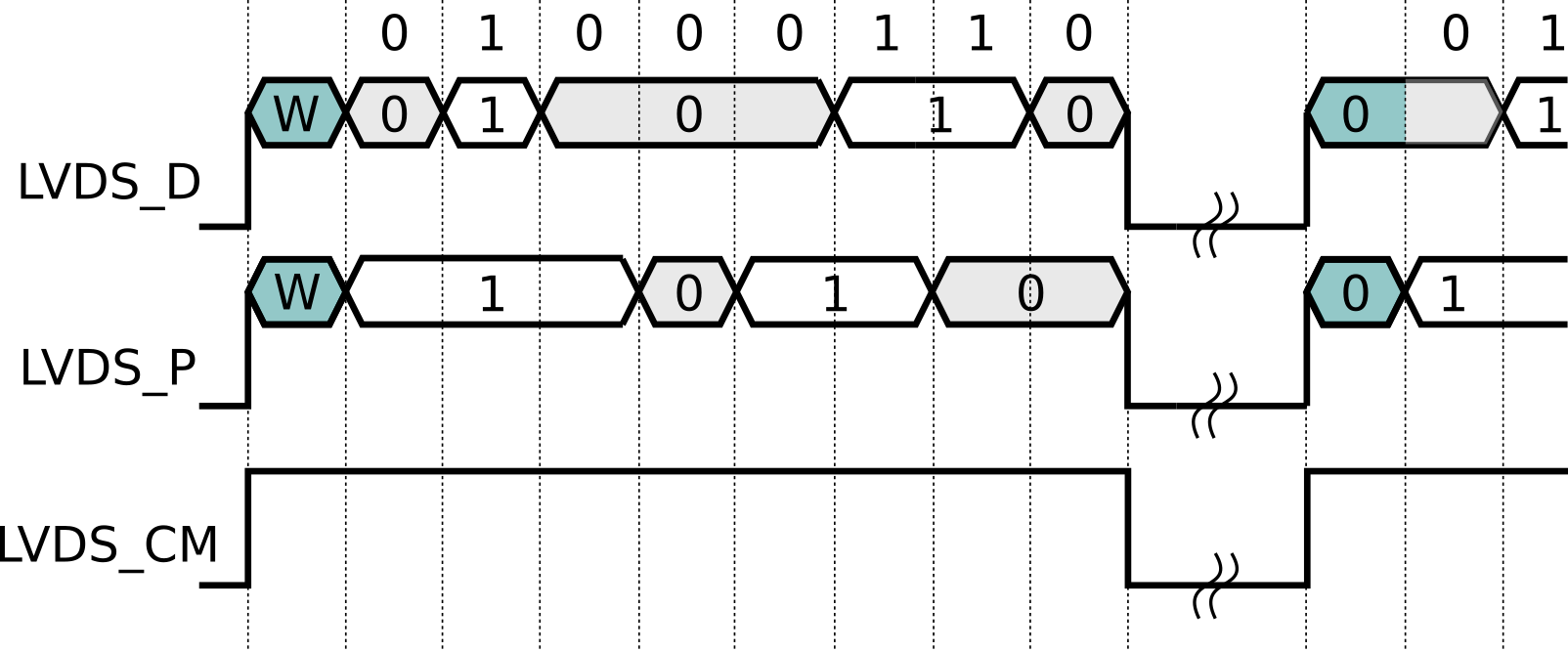}
  \caption{Proposed signaling scheme with LVDS for data transmission and common-mode voltage for instant on/off receiver. }
  \label{fig:LVDS_CM}
\end{figure}

Instant on/off \ac{LVDS} drivers and receivers that implement event-driven wake-up and sleep-mode mechanisms are crucial for minimizing consumption in neuromorphic systems that operate with sparse activity and low average event rates. Since the main digital blocks communicate with each other following a four-phase handshaking protocol, no dynamic power is dissipated in idle states. The ``\ac{LVDS} Drivers''  can be easily turned on or off by a digital signal, such as $TX.r$ of Fig.~\ref{fig:arch}, as new event data appears on the ``Input Buffer'' block. In order to turn on/off the ``\ac{LVDS} Drivers''  instantly, we exploited the common voltage of \ac{LVDS} pairs.  As shown in Fig.~\ref{fig:LVDS_CM}, during the idle state, when no data is being transmitted, the two pairs of \ac{LVDS} signals are both pulled down to \textsl{Gnd}, resulting in a 0V common mode voltage. In this way the ``\ac{LVDS} Receivers'' , designed with NMOS input transistors, will be fully tuned off and power consumption will be due only to off-leakage level static power dissipation. As soon as a new event arrives, the common-mode feedback circuit will drive both data pair and parity pair voltage lines back to a \textsl{Vref} common mode voltage, which is set to about 1V in this design. Simultaneously, the differential voltages of data pair and parity pair will recover back to their previous bits value with $D=P$. In this way the ``\ac{LVDS} Receivers''  with NMOS input transistors will be turned on and will start to convert the \ac{LVDS} signals to standard digital ones. Because the first odd token-cell in the decoder will only take data when $P[i]=\overline{D[i]}$, the receiver will ignore potential spurious repeated  LSB bits until a new MSB bit arrives. After transmitting the full word, the common-mode voltage of the \ac{LVDS} pairs will be pull down to \textsl{Gnd} again, turning the receiver off. The recovery speed of common-mode voltage is controlled by a common-mode feedback circuit in the \ac{LVDS} driver. In our measurement, the recovery latency of common-mode voltage is less than 0.5\,ns, which is much shorter than previously reported values (e.g., 6.6\,ns in~\cite{Zamarreno-Ramos_etal12}).

\subsection{Transmission Scheme}
\begin{figure}
  \centering
  \includegraphics[width=0.48\textwidth]{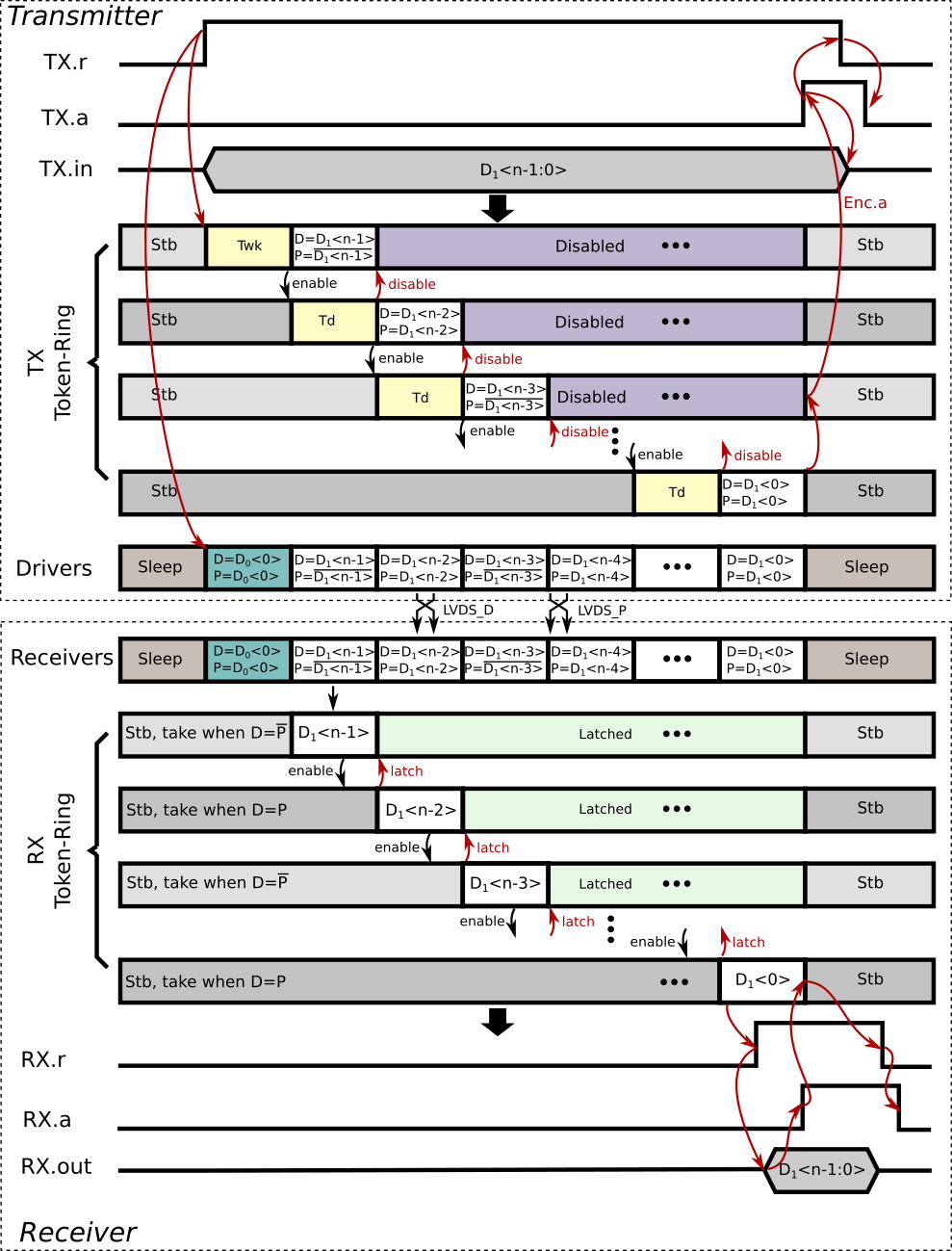}
  \caption{Event transmission timing diagram of the bit-serial \ac{LVDS} link. The ``Stb'' represents the stand-by state.}
  \label{fig:scheme}
\end{figure}

Figure~\ref{fig:scheme} describes the timing diagram for the transmission of one event in the proposed bit-serial \ac{LVDS} link. A four-phase handshaking protocol is implemented between ``Input Buffer'' and ``TX Token-Ring''. Once the event data $D_{1}<n-1:0>$ appears on the ``TX Token-Ring'' input bus $TX.in$, the signal $TX.r$ will be set to high by the ``Input Buffer'', thus requesting a new data transmission which will trigger the first stage token-cell of ``TX Token-Ring'' to take the first bit. Meanwhile, this request signal will turn on the \ac{LVDS} drivers to be ready for sending new data. After a tunable delay $t_{wd}$, the first odd token-cell of ``Token-Ring'' will push new bit value $D=D_{1}<n-1>$ and parity $P=\overline{D_{1}<n-1>}$ to shared data and parity wires $Data$ and $Parity$. It will then enable the following stage, i.e., the first even token-cell to take new bit value. After a cycle delay $t_{d}$, the enabled stage will disable previous stage and push new data/parity $D=D_{1}<n-2>$ and $P=D$ on the shared wires, while enabling its following stage. Mutual exclusion is implemented stage by stage till the end the token-ring. After pushing data/parity of the last bit value to shared wires, the ``TX Token-Ring'' block will acknowledge the ``Input Buffer''  by asserting $Enc.a$ to high. Subsequently, $TX.r$ will be reset to low for the successful removal of data $D_{1}<n-1:0>$. Finally, $TX.a$ will be reset to low to complete the four-phase handshaking cycle. 

It should be noted that during the wake-up stage the ``\ac{LVDS} Receivers''  will need to be turned on for recovering the common-mode voltage of \ac{LVDS} pairs with data and parity value $P=D$. The first token-cell of ``RX Toke-Ring'' will only take data and parity with $P=\overline{D}$. For an event data with even bit width, a safe approach is to fully recover both common-mode and differential values of previous bit by repeating the LSB of the previous event data with data $D=D_{0}<0>$ and party $P=D_{0}<0>$. 

The mutually exclusive token-cells of ``RX Token-Ring'' will take data from \ac{LVDS} receivers bit-by-bit. Each bit cycle is distinguished by either $P=\overline{D}$ or $P=D$. The output of each token-cell is latched, once the current token-cell is disabled by its successor. As soon as the last token-cell gets its bit, it will request ``Output Buffer'' to take the whole data packet from all token-cells and reset the ``RX Token-Ring''. In this design the ``RX Token-Ring'' is required to have the highest throughput. The tunable delay $T_{d}$ is  added in ``TX Token-Ring'' to enforce the timing assumptions that ``RX Token-Ring'' has a higher throughput than ``TX Token-Ring'', to get sequence bit within one TX bit cycle.

\section{Circuits Implementation}
\label{sec:circ-impl}

\subsection{Transmitter Token-Ring}

\begin{figure}
  \centering
  \includegraphics[width=0.44\textwidth]{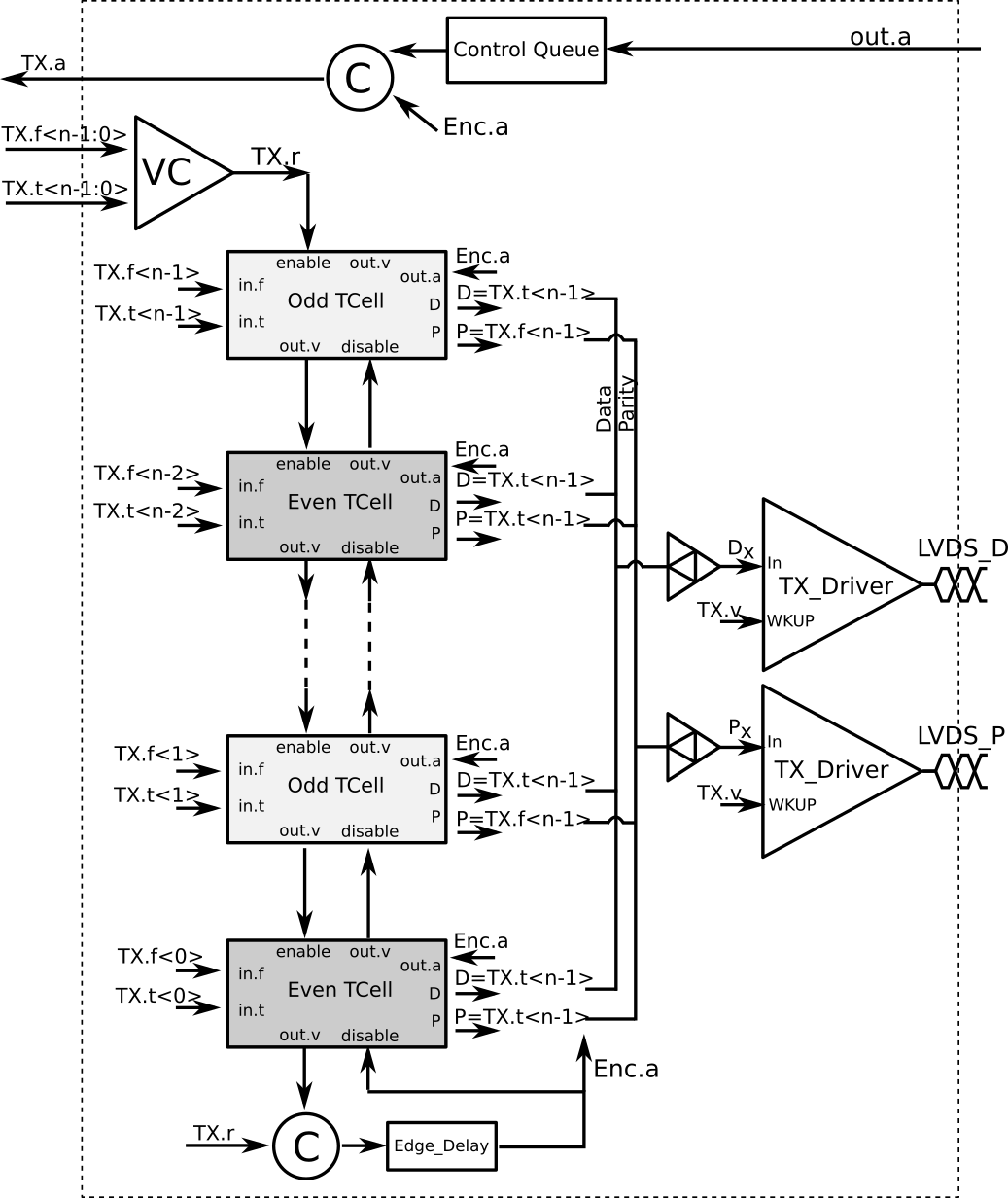}
  \caption{Transmitter Token-Ring for encoding data into data/phase scheme in proposed bit-serial \ac{LVDS} link. }
  \label{fig:TX_token_ring}
\end{figure}

\begin{figure}
  \centering
  \includegraphics[width=0.48\textwidth]{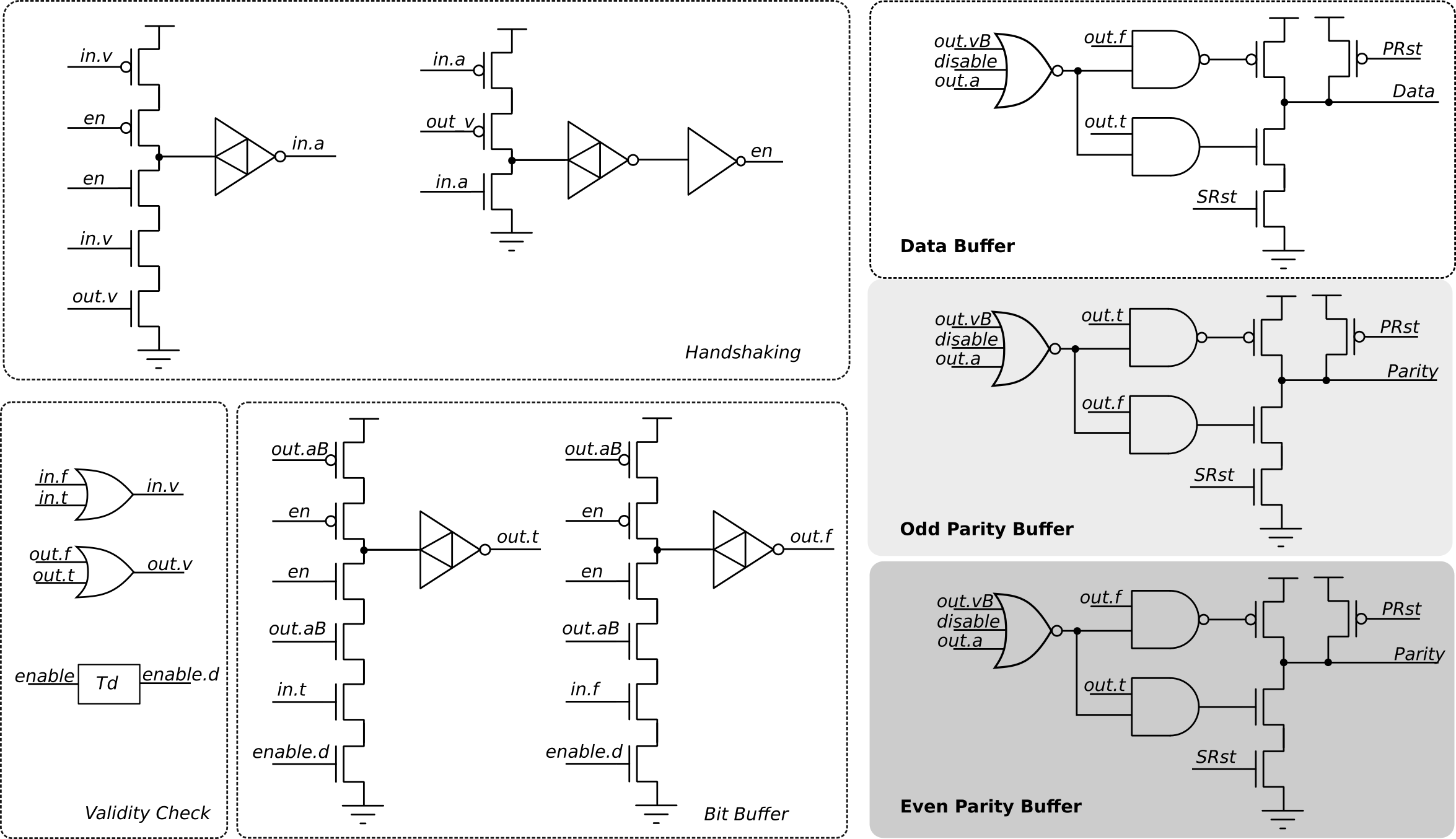}
  \caption{Circuit implementation of the TX token-cell based on a bit-buffer. Each token cell comprises ``Handshaking'', ``Validity Check'', ``Bit Buffer'', ``Data Buffer'' and ``Odd/Even Parity Buffer'' blocks.  }
  \label{fig:TX_token_cell}
\end{figure}

The block diagram of ``TX Token-Ring'' is shown in Fig.~\ref{fig:TX_token_ring}. A dual-rail asynchronous protocol and four-phase handshaking are used for processing input data. The ``TX Token-Ring'' comprises an input ``Validity Check'' block, ``Token-Ring'' with odd and even token-cells, ``\ac{LVDS} Drivers'' and a ``Control Queue'' block.
The ``Validity Check'' block first checks and indicates a valid input event data by $TX.r$. The ``\ac{LVDS} Drivers'' can then be turned on by $TX.r$ for a valid input event. Meanwhile, the first token-cell starts to take the first bit value $TX.f<n-1>/TX.t<n-1>$ and push relative data and parity outputs to shared wires. For odd bits, data and parity outputs are $D=B$ and $P=\overline{D}$, respectively,  while for even bits, they are $D=B$ and $P=D$, respectively. So the first odd token-cell will push $D=TX.t<n-1>$ to shared data wire and $P=TX.f<n-1>$ to shared parity wire. After a tunable delay $t_{wk}$ when the first token-cell successfully pushes data and parity value of MSB of input event to shared data and parity wires, the first token-cell will send the $enable$ signal to enable its successor for the next bit value. As a response, its successor will send back the $disable$ signal as soon as it successfully takes a bit value. After a set ``bit cycle'' time $t_{d}$ when the last token-cell pushes its output to shared data and parity wires, $Enc.a$ will be asserted to high to reset the whole ``TX Token-Ring'' and acknowledge ``Input Buffer'' to erase old data, and will be reset to low to acknowledge that old data has returned to zero ($TX.f<n-1:0>=0, TX.t<n-1:0>=0$). At this point the ``TX Token-Ring'' is free to take new data.


Figure~\ref{fig:TX_token_cell} shows the circuit implementation of the proposed token-cell, based on an asynchronous buffer following a dual-rail protocol and four-phase handshaking. The token-cell comprises a ``Handshaking'' block, a ``Validity Check'' block, ``Bit Buffer'', ``Data Buffer'' and ``Odd/Even Parity Buffer'' blocks. The ``Validity Check'' block checks the validity of input bit value and indicates the state by signal $in.v$. The ``Handshaking'' block generates the acknowledge signal $in.a$ to acknowledge a valid bit input and control signal $en$ to enable Bit Buffer block for buffering current input bit value ($en=1$) or reset the ``Bit Buffer'' block for the next cycle ($en=0$). The ``Data Buffer'' and ``Odd/Even Parity Buffer'' blocks will convert buffered bit value $out.t$ and $out.f$ to data and parity value according to \ac{LEDR} protocol and push them to shared $Data$ and $Parity$ wires. Once the current token-ring generates a valid output which is indicated by $out.v=1$, it will enable its successor and disable its predecessor for mutual exclusion.

\subsection{LVDS Drivers}

\begin{figure}
  \centering
  \includegraphics[width=0.4\textwidth]{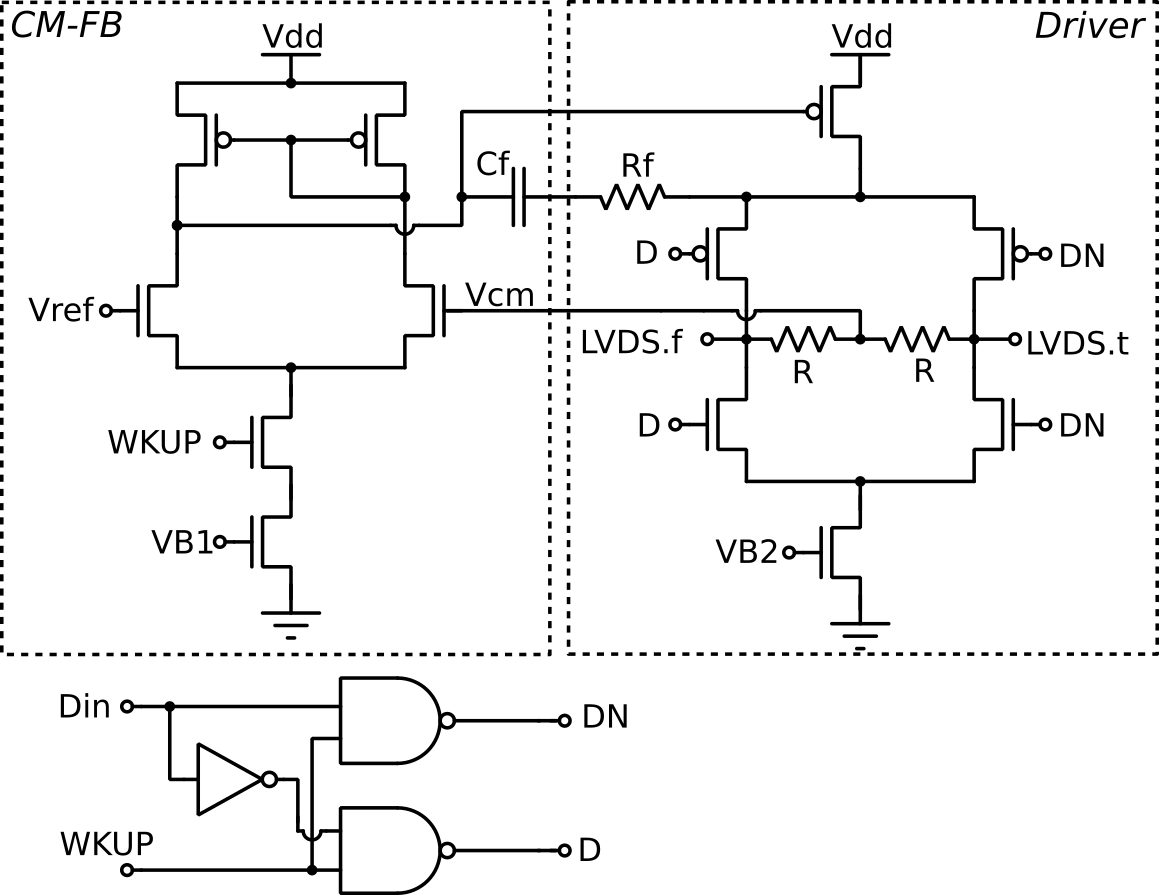}
  \caption{Circuit implementation of the ``TX \ac{LVDS} Driver''. }
  \label{fig:TX_driver}
\end{figure}

Current mode \ac{LVDS} Drivers, shown in Fig.~\ref{fig:TX_driver}, are used to convert data and parity value on shared wires to \ac{LVDS} pairs. The ``\ac{LVDS} Driver'' is implemented such that it can convert input value $D_{in}$ into \ac{LVDS} signals for a valid input ($WKUP=1$) and fully turned off for a standby mode ($WKUP=0$). When no data is transmitted ($WKUP=0$), the ``CM-FB'' block is switched off. The signals $DN$ and $D$ are then both set to logic "1" to tune off their gating PMOS transistors and tune on their gating NMOS to pull both $LVDS\_f$ and $LVDS\_t$ down to \textsl{Gnd}, with a common-mode \ac{LVDS} pair voltage $V_{CM}=0$. This will switch off its linked ``\ac{LVDS} Receiver'' block following the proposed instant on/off scheme. Once there is a valid input ($WKUP=1$), the ``CM-FB'' block will supply property common-mode \ac{LVDS} pair voltage $V_{CM}=V_{ref}$ to switch on the ``\ac{LVDS} Receiver'' on the receiver side, and the ``Driver'' block will start to convert $D_{in}$ into \ac{LVDS}. \revised{}{The $V_{B1}$ and $V_{B2}$ signals are biases to generate proper tail currents for the ``CM-FB'' and ``Driver'' blocks.}{} Two resistors with value $R=50\Omega$ (with another two resistors placed at the input terminals of the ``\ac{LVDS} Receiver'') are used to setup differential amplitude of \ac{LVDS} pair.

\subsection{Receiver Token-Ring}
\begin{figure}
  \centering
  \includegraphics[width=0.44\textwidth]{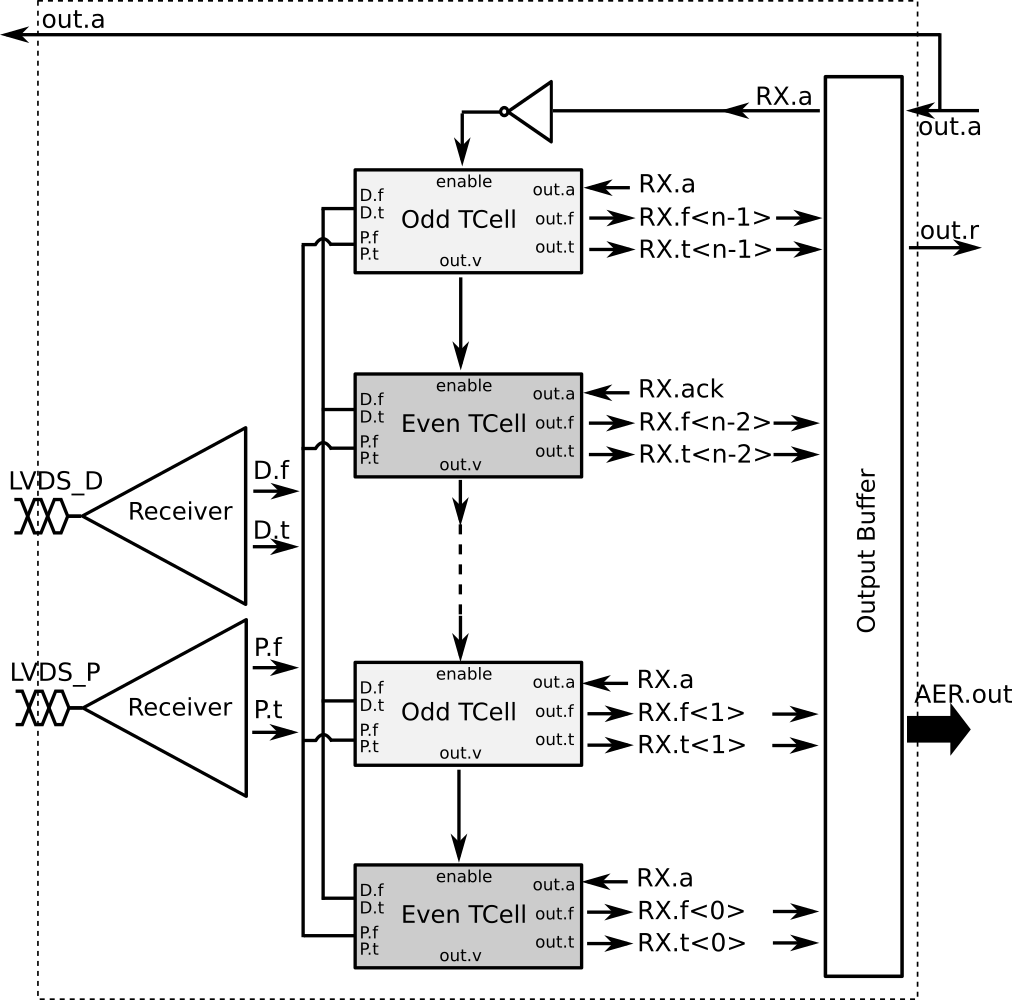}
  \caption{Receiver Token-Ring for decoding data/phase to event data in proposed bit-serial \ac{LVDS} link.  }
  \label{fig:RX_token_ring}
\end{figure}

\begin{figure}
  \centering
  \includegraphics[width=0.44\textwidth]{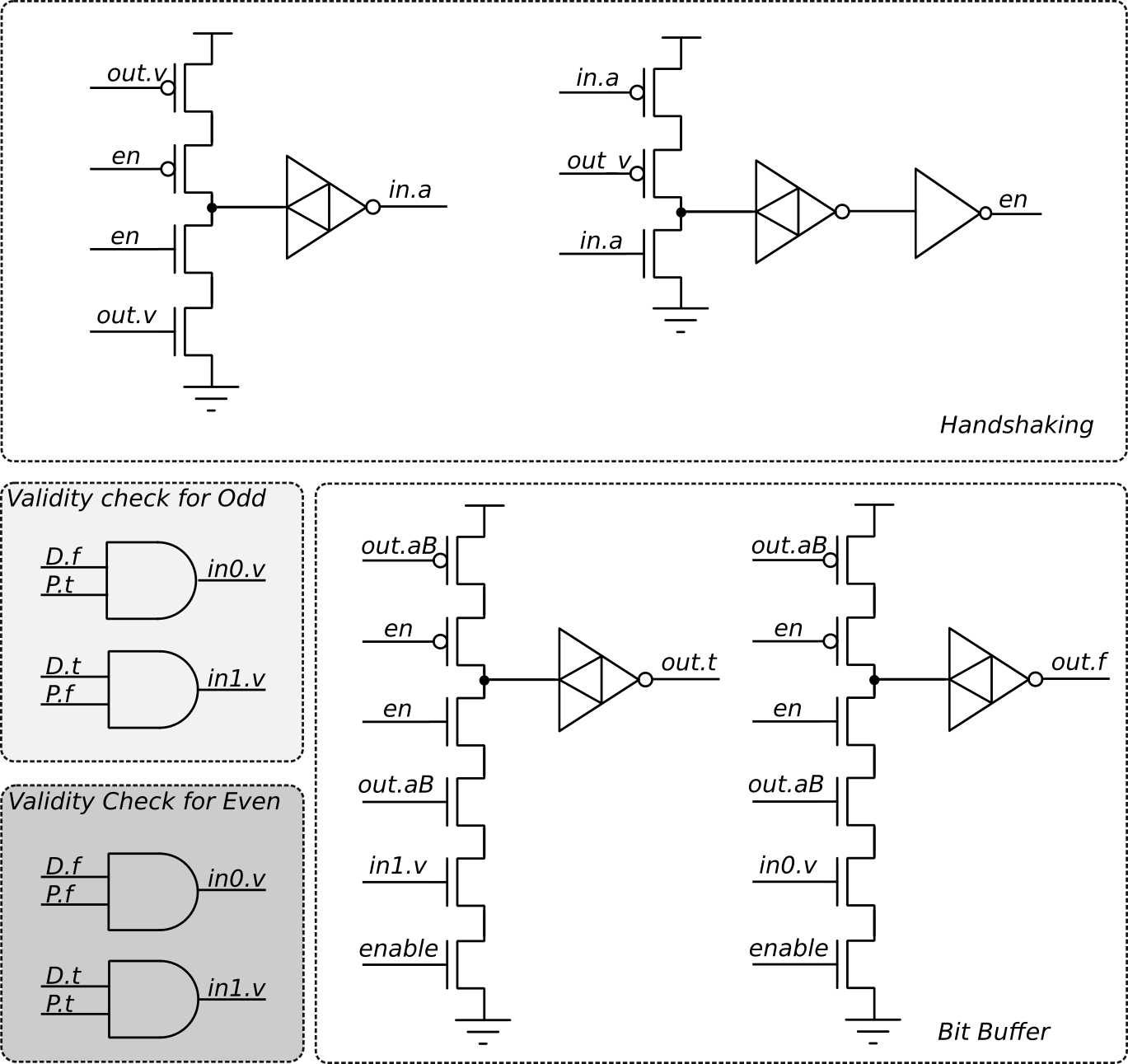}
  \caption{Circuit implementation of RX token-cell based on 1-bit buffer. Each token-cell comprises Handshaking block, Validity Check block and Bit Buffer block. }
  \label{fig:RX_Tcell}
\end{figure}

The architecture of the ``RX Token-Ring'' for processing and decoding \ac{LVDS} pairs $LVDS\_D$ and $LVDS\_P$ is shown in Fig.~\ref{fig:RX_token_ring}. Following the dual-rail asynchronous protocol and four-phase handshaking, the ``RX Token-Ring'' comprises ``\ac{LVDS} Receivers'', ``Token-Ring'' with odd and even token-cells and an ``Output Buffer'' block. 
The ``\ac{LVDS} Receivers'' first digitize the \ac{LVDS} pairs $LVDS\_D/P$ to digital sequential bits $D.f/t$ and $P.f/t$, respectively. The token-cells then take bits one-by-one till the end of this event transmission. Once all token-cells take and buffer bits value, the following the ``Output buffer'' will buffer received event data $RX.f<n-1:0>$ and $RX.t<n-1:0>$ to output bus $AER.out$ and reset the ``RX Token-Ring'' for new data.

The circuit implementation of the ``RX token-cell'' following the dual-rail protocol and four-phase handshaking is shown in Fig.~\ref{fig:RX_Tcell}. Each RX token-cell comprises a ``Handshaking'', ``Validity Check'' and ``Bit Buffer'' block. The odd token-cell will only take and process bit value with $P=\overline{D}$ and even token-cell will only take and process bit value with $P=D$. When new bit value comes to the token-ring with proper data and parity value relationship, for example, $D.f=P.t$ and $D.t=P.f$ for $P=\overline{D}$, the current activated odd token-cell will take this bit value and buffer it with its ``Bit Buffer'' block. After generating a valid output bit value ($out.v=1$), internal signal $en$ will be set to logic ``0'' to latch the output bit value and block it to take new bit value. Meanwhile, this token-cell will enable its following token-cell for a new token.

\subsection{LVDS Receivers}

\begin{figure}
  \centering
  \includegraphics[width=0.44\textwidth]{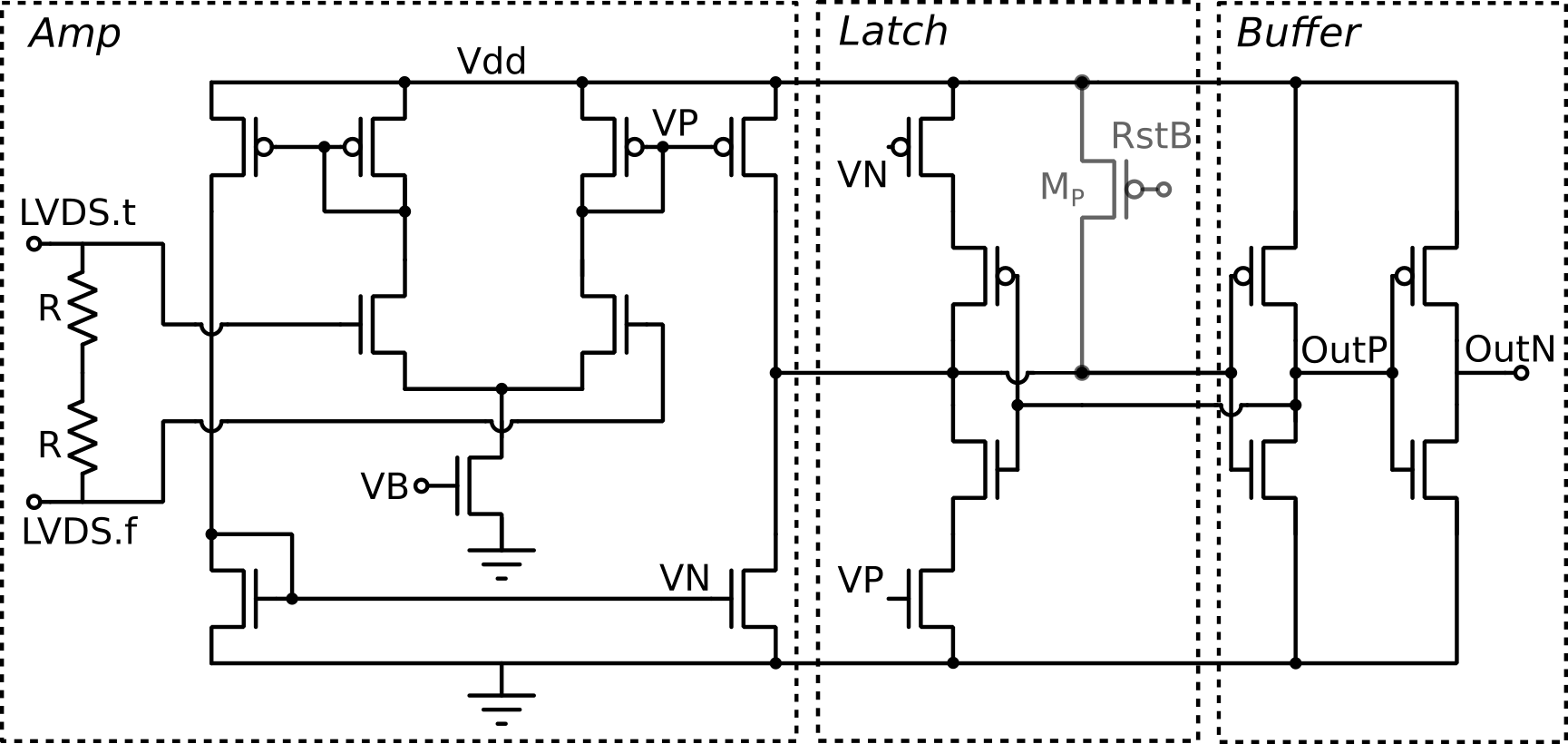}
  \caption{LVDS Receiver for digitizing differential \ac{LVDS} to digital signals. }
  \label{fig:RX_amp}
\end{figure}
In order to meet the requirement of instant on/off by means of the \ac{LVDS} common-mode voltage, we implemented the amplifier-based \ac{LVDS} receivers with NMOS inputs. The circuit implementation of the proposed ``LVDS Receiver'' is shown in Fig.~\ref{fig:RX_amp}. It comprises an ``Amp'' block, a ``Latch'' block and a ``Buffer'' block. The ``Amp'' block is responsible for digitizing the \ac{LVDS} signals. In standby mode, the ``Amp'' stage will be fully tuned off with $LVDS\_f/t=0$. Once there is data from transmitter side that needs to be transmitted  (i.e., when $V_{CM}=V_{ref}$), the ``Amp'' stage will be tuned on instantly. A latch stage with dynamic biases is implemented to latch the last bit value of previous event data once the ``Amp'' stage is switched to sleep mode so that the \ac{LVDS} receiver will not wake up with a random output bit value. After a successful event transmission, $V_{CM}$ of the \ac{LVDS} pair will be switched from $Vref$ to \textsl{Gnd},  with $VP$ and $VN$ shifting to \textsl{Vdd}  and \textsl{Gnd} respectively. This will strengthen the drive ability of the latch stage to store the current bit value when the ``Amp'' stage is turned off. As new event data arrives, the signals $VP$ and $VN$ will be shifted to near \textsl{Vdd/2} to tune the latch stage weaker, so that it can be modified by the new data. An active-low reset signal $RstB$ is used to reset the circuit outputs to a proper initial condition ($P=D$) when powering up the chip.

\section{Experimental results}
\label{sec:experiment}
\begin{figure}
  \centering
  \includegraphics[width=0.42\textwidth]{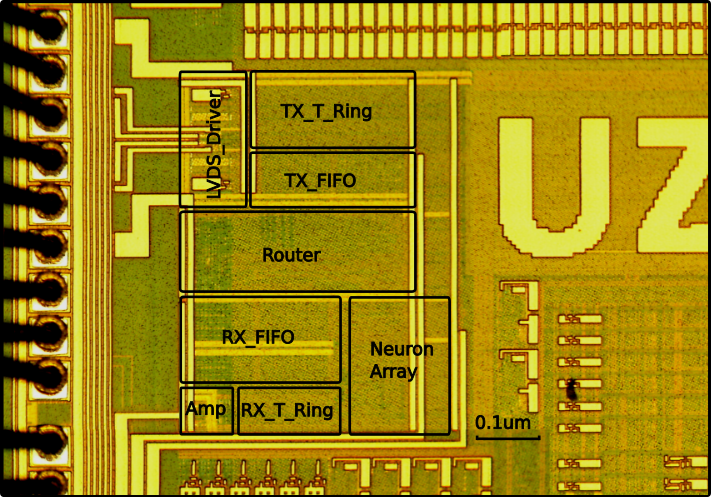}
  \caption{Die photo of test chip with proposed event-driven bit-serial \ac{LVDS} link in AMS 0.18\,um 1P6M process, in which TX block occupies an area of 0.08 \,$mm^{2}$ and RX block occupies an area of 0.06\,$mm^{2}$.}
  \label{fig:die_photo}
\end{figure}

\begin{figure}
  \centering
  \includegraphics[width=0.4\textwidth]{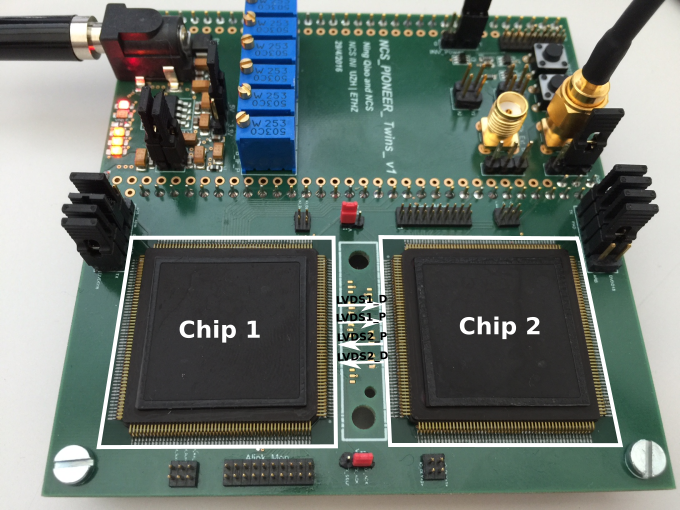}
  \caption{The setup for testing \ac{LVDS} links between two chips for bidirectional communication. }
  \label{fig:setup}
\end{figure}

The proposed fully asynchronous event-driven bit-serial \ac{LVDS} link was implemented using a standard 0.18\,$\mu$m 1P6M CMOS process, occupying a silicon area of 0.14\,$mm^{2}$. Figure~\ref{fig:die_photo} shows the die photo of the fabricated test chip. The whole ``Transmitter'' block including the ``TX\_Buffer'' occupies an area of 0.08 \,$mm^{2}$, and the ``Receiver'' block including the ``RX\_Buffer'' occupies an area of 0.06\,$mm^{2}$. Additionally, a small spiking neural array with tunable output event rate is implemented to provide events for testing. A 32-bit router is implemented for routing events from Receiver to Transmitter to realize a transmission loop between 2 chips to explore peak transmission throughput.

\begin{figure}
  \centering
  \includegraphics[width=0.46\textwidth]{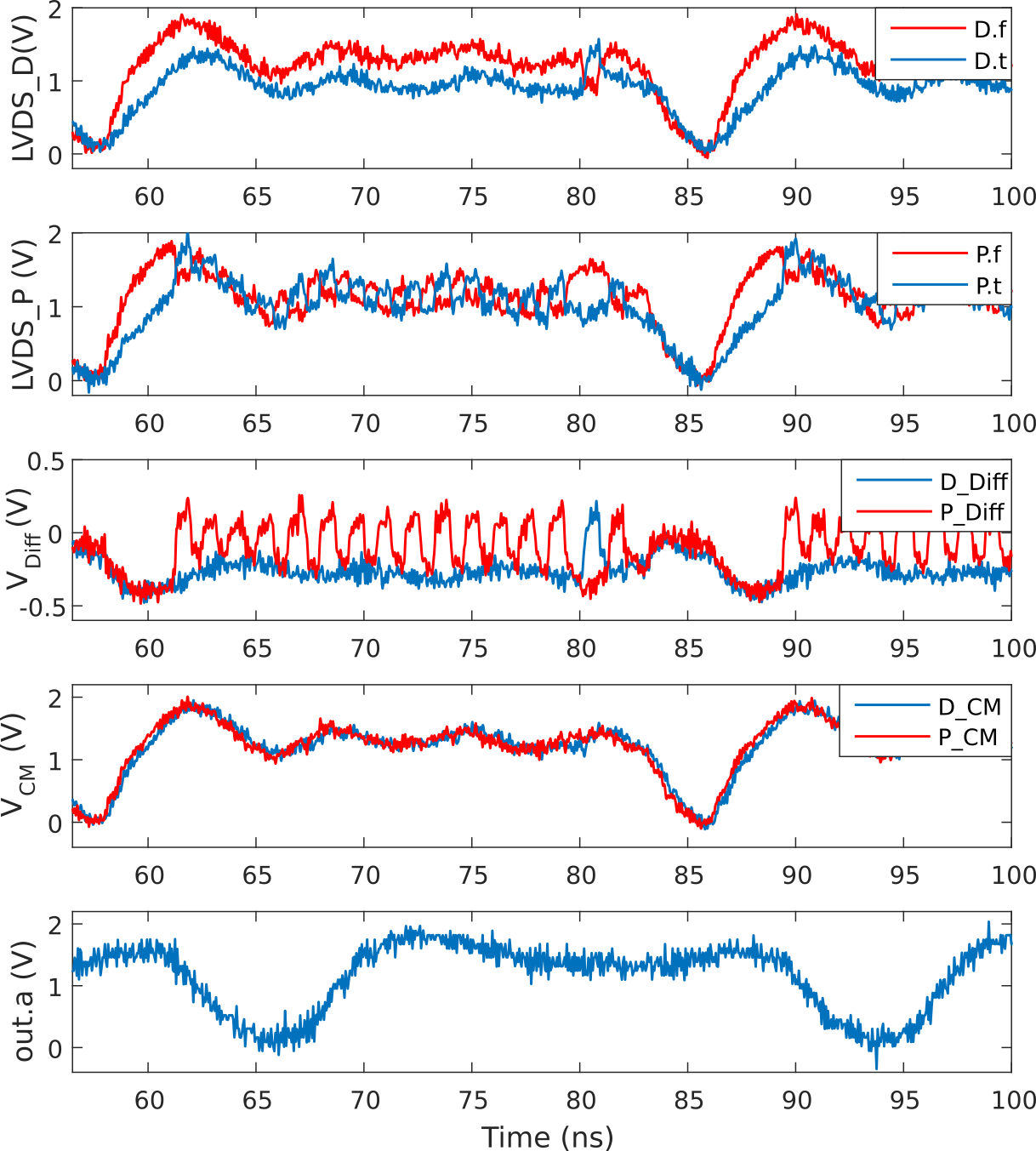}
  \caption{Transient signals of \ac{LVDS} pairs captured on the receiver's inputs: the traces $D.f$ and $D.t$ represent the differential signals for $LVDS\_D$; the traces $P.f$ and $P.t$ represent the differential signals for $LVDS\_P$; The $D\_Diff$ and $P\_Diff$ traces are differential voltages of two \ac{LVDS} pairs; The $D\_CM$ and $P\_CM$ traces represent the common voltages of the two \ac{LVDS} pairs; The last plot shows the $RX\_Ack$ signal, which is the acknowledge signal from the target chip to the source chip, representing a successful event transmission.}
  \label{fig:experiment}
\end{figure}

\begin{figure}
  \centering
  \includegraphics[width=0.46\textwidth]{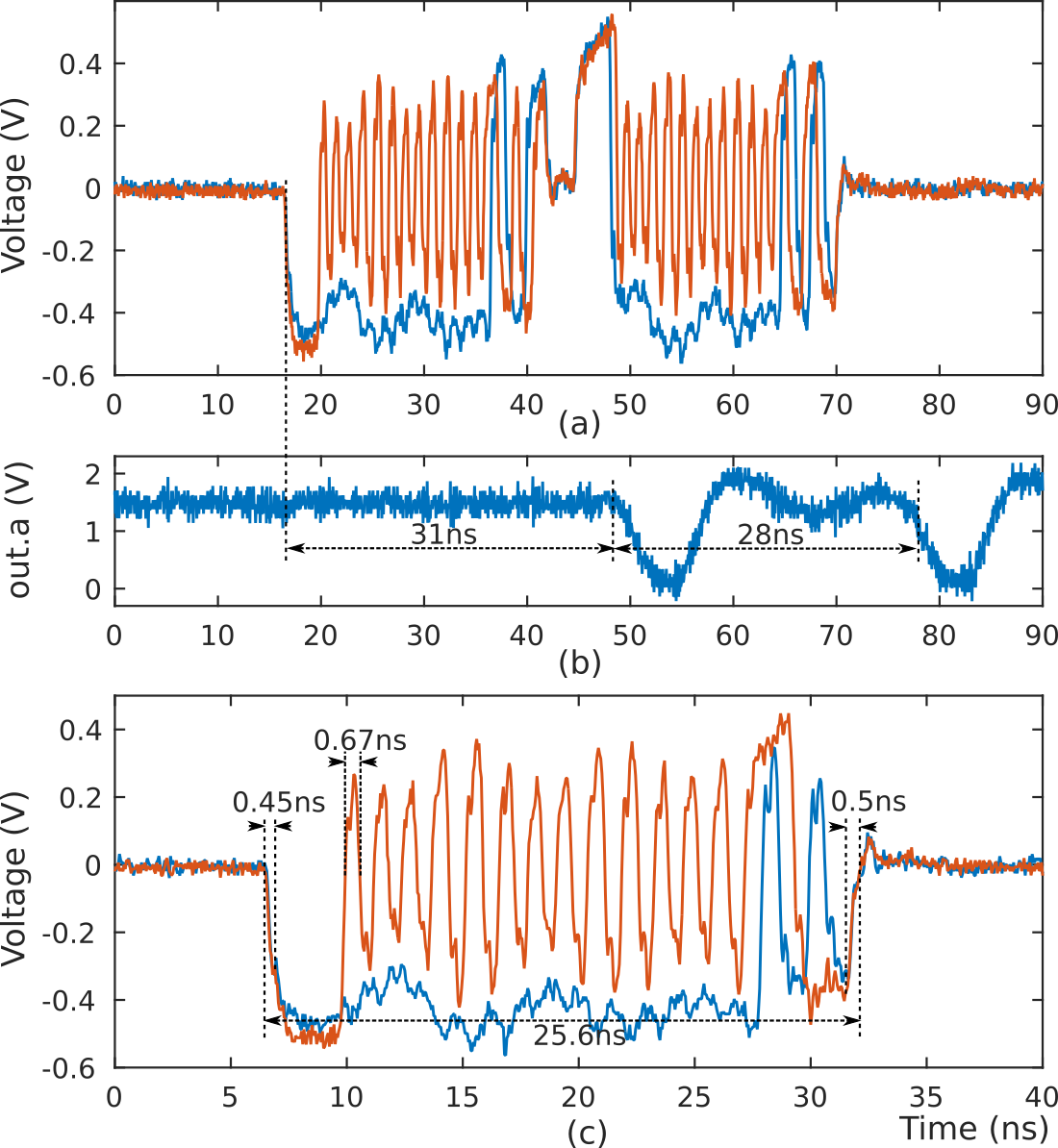}
  \caption{Transient signals of \ac{LVDS} pairs at receiver inputs: (a) differential mode of \ac{LVDS} signals, (b) acknowledge signal from the receiver, (c) details of single event transmission signals. }
  \label{fig:experiment2}
\end{figure}

Figure~\ref{fig:setup} shows a setup with two chips placed side-by-side for the experiments. With this setup, we transmitted sequences of 32-bit AER events  bi-directionally between two chips, through four \ac{LVDS} pairs: The signals $LVDS1\_D$ and $LVDS 1\_P$ were used to transmit events from Chip1 to Chip2, and  $LVDS2\_D$ and $LVDS 2\_P$ were used to transmit events from Chip2 to Chip1.

Transient Signals of \ac{LVDS} pairs were observed and captured using a Tektronix DPO7000 Oscilloscope, from the input terminals of the ``\ac{LVDS} Receivers''. As shown in Fig.~\ref{fig:experiment}, the $LVDS\_D$ plot shows data from the \ac{LVDS} pair with differential signals $D.f$ and $D.t$. The  $LVDS\_P$ plot shows the parity \ac{LVDS} pair with differential signals $P.f$ and $P.t$. The $D\_Diff$ and $P\_Diff$ traces in the $V_{Diff}$ plot are the differential voltages of the data \ac{LVDS}  and parity \ac{LVDS} pairs, respectively. The $D\_CM$ and $P\_CM$ traces in the $V_{CM}$ plot are the common voltage of data \ac{LVDS} and parity \ac{LVDS} pairs, respectively. The $out.a$ plot shows the acknowledge signal from the target receiver chip for acknowledging a successfully event transmission. Sequence bits are presented bit-by-bit following the \ac{LEDR} protocol, via the data and parity differential signals $D\_Diff$ and $P\_Diff$. The common-voltages of the two pairs $D\_CM$ and $P\_CM$ are reset to \textsl{Gnd} at the end of a successful event transmission and are quickly recovered with new coming events. During the recovery of common-mode voltages of the \ac{LVDS} pairs, the LSB of previous event with $P=D$ is repeated for sufficient long time to guarantee that the receiver is fully and successfully switched on.

\begin{table*}
  \caption{Performance comparison of \ac{LVDS} transceiver VLSI implementations.}
  \centering
  \begin{tabular}{l  c  c  c c}
     \toprule
         & \textbf{\cite{Teifel_Manohar03}} & \textbf{\cite{Zamarreno-Ramos_etal08}} & \textbf{\cite{Zamarreno-Ramos_etal13a}} & \textbf{This work}\\
    \midrule
    \textbf{Technology} & 0.18\,$\mu$m & 90\,nm & 0.35\,$\mu$m & 0.18\,$\mu$m \\

    \textbf{Power Supply}  & 1.8\,V & 1\,V  & 3.3\,V & 1.8\,V \\

    \textbf{Area}  & 0.016\,$mm^{2}$ & 0.09\,$mm^{2}$  & 0.352\,$mm^{2}$ & 0.14\,$mm^{2}$  \\

    \textbf{Clocked CDR} & No & Yes   & Yes & No   \\   

     \textbf{Bit Rate} & 3\,Gps& 1\,Gps  & 0.64\,Gps & 1.5\,Gps  \\

     \textbf{Event Rate} & - & 29.4\,M\,Event/s & 13.7\,M\,Event/s  & 35.7\,M\,Event/s  \\

    $\mathbf{P_{max}}$ & 77\,mA  & 40.1\,mA  & 15.9\,mA & 22.9\,mA   \\

    $\mathbf{P_{min}}$  & - & 40.1\,mA  & 0.4\,mA & 0.122\,$\mu$A  \\       

    $\mathbf{P_{max}/P_{min}}$  & - & 1 & 39 & 187.7k \\       
    \bottomrule
  \end{tabular} 
  \label{table:performance} 
\end{table*}

Figure~\ref{fig:experiment2} shows the transient signals of the \ac{LVDS} pairs at the input terminals of receiver chip, captured by an \ac{LVDS} probe TektronixP6880. The bit cycle is set to be around 0.67\,ns by tuning the delay cell $t_{d}$ in the ``TX Token-Ring'' to achieve a bit-rate of 1.5\,Gps. The observed switch on/off speed of the receiver is approximately 0.45\,ns and 0.5\,ns, respectively, leading to a smaller latency for \ac{AE} transmission. As measured in Fig.~\ref{fig:experiment2}(b), the latency needed for a successful transmission between chips (from switching on the Receiver to getting acknowledge signal $out.a$ from receiver chip) for a 1.5\,Gps bit-rate is 31\,ns. The period of a successive events transmission is 28\,ns. Since the transmitter has locally pre-stored the ``out.a'' signal in the ``Control Queue'' block (see Section~\ref{sec:token_ring}),  it will keep on sending event data without waiting for acknowledge signals from the receiver chip until the ``Control Queue'' is fully empty, to further decrease latency. This is evidenced in Fig.~\ref{fig:experiment2}(a) and (b), as the second event transmission happens before the arrival of first acknowledge signal $out.a$. In  Fig.~\ref{fig:experiment2}(c) we can observe that, for transmitting a 32-bit event data with a bit-rate of 1.5\,Gps, the \ac{LVDS} link will only be switched on for 25.6\,ns, and will be switched off instantly on both transmitter and receiver sides, leading to a pure event-rate related power consumption. 

\begin{figure}
  \centering
  \includegraphics[width=0.46\textwidth]{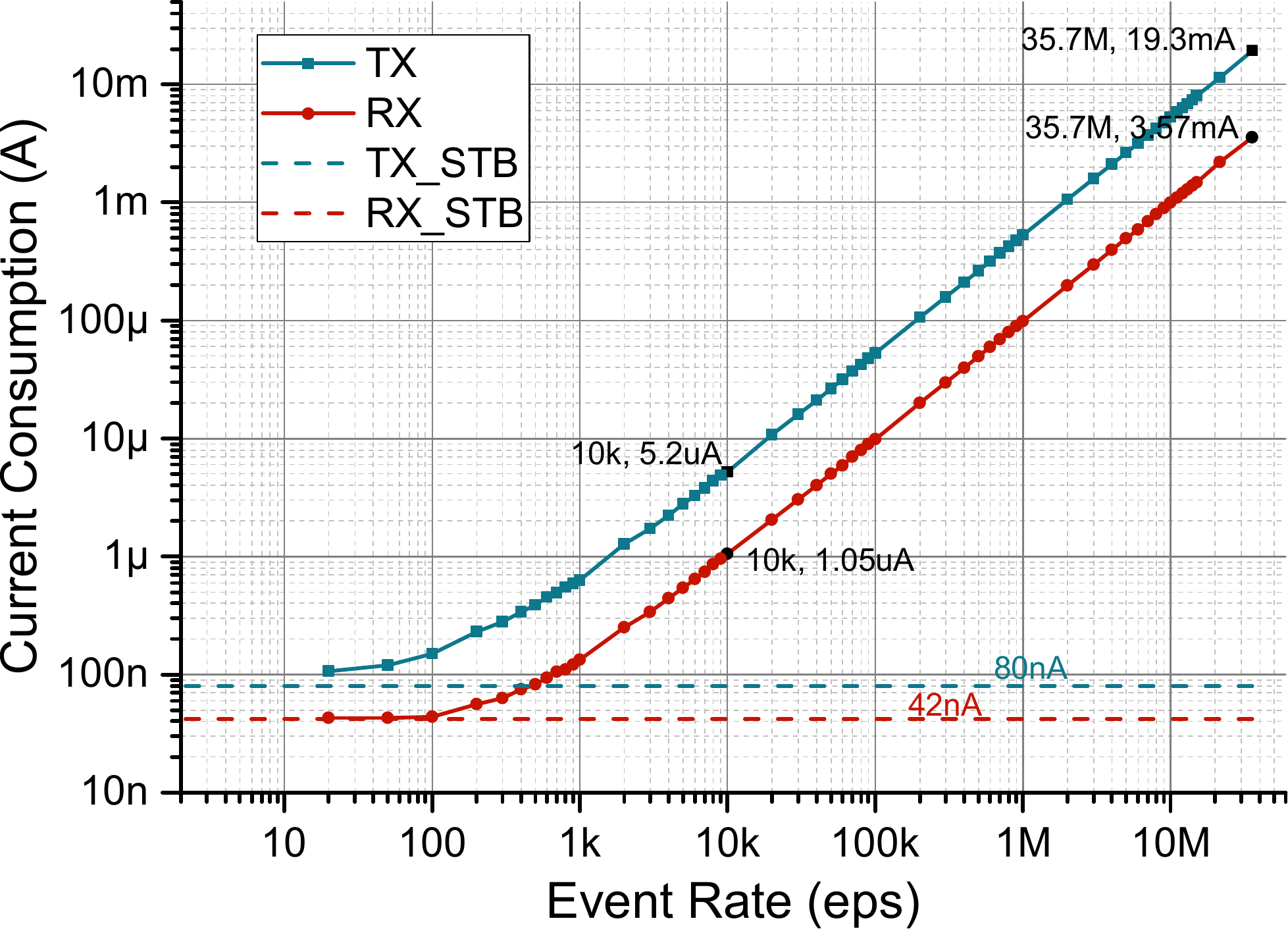}
  \caption{Power consumption of asynchronous serial-bit \ac{LVDS} link. }
  \label{fig:power}
\end{figure}

In Fig.~\ref{fig:power} we plot the measured power consumption for different event transmission rates. The peak event rate that can be achieved in our experimental setup is 35.7\,M\,Events/second (32-bit) with current consumption of 19.3\,mA and 3.57\,mA for transmitter and receiver part, respectively. The power consumption of both transmitter and receiver part scales linearly with the event transmission rate. At a 10k event rate, the power consumption of the transmitter and receiver blocks are 5.2\,$\mu$A and 1.05\,$\mu$A, respectively. The power consumption can further go down to sub-$\mu$A for a lower event rates (<1k\,Events/second), with a floor of 80\,nA for transmitter and 42\,nA for receiver which is mainly dominated by leakage current of circuits. 

Table~\ref{table:performance} shows a performance comparison between different designs. However, area and power consumption of \ac{CDR} circuits employed in the designs of~\cite{Zamarreno-Ramos_etal08, Zamarreno-Ramos_etal13a} are not reported. So it may be that significant additional silicon area and power consumption are required for those designs.


\section{Conclusions}
\label{sec:conclusions}

While neuromorphic electronic systems have the potential of solving the memory bottleneck problem~\cite{Indiveri_Liu15}, by construction they also face an important I/O bottleneck problem: large scale neuromorphic system implementations are typically composed of multiple cores and/or multiple chips tiled together, with grid-like communication networks. To transmit address-events across these cores and chips and to sustain the required bandwidth, current implementations use multiple parallel \ac{AER} buses (e.g., for North-South, East-West, and possibly diagonal links). In this paper we argued that full parallel or even word-serial \ac{AER} protocols are not scalable, as they require large number of pins/pads and large power consumption to quickly charge and discharge all these lines. To solve this problem, we proposed an ultra low-power fully asynchronous event-driven instant on/off bit-serial \ac{LVDS} link, which is suitable for \ac{AER} transmission in neuromorphic multi-chip systems. The proposed \ac{LVDS} link uses \ac{LEDR} encoding and a token-ring architecture to eliminate the need for clock-based \ac{CDR} blocks with expensive on chip DLL/PLL circuits, leading to a very compact and low-power circuit implementation. A novel scheme is proposed to implement a low-latency event-driven transmission with sub-ns instant on/off feature. Experimental results demonstrate how the proposed bit-serial \ac{LVDS} link can achieve an event rate of 35.7\,M\,Events/second with a bit-rate of 1.5\,Gps. The power consumption of the proposed \ac{LVDS} link is pure rate-dependent, with a sub-$\mu$A  power consumption for low event rates (e.g.,$\approx$1k\,Events/second).

\section*{Acknowledgment}
This work is supported by the EU ERC 
grant ``NeuroP'' (257219) and by the EU ICT grant ``NeuRAM$^3$'' (687299). 

\bibliographystyle{IEEEtran}


\end{document}

%% file: acronyms.tex
\acrodefplural{RAM}[RAMs]{Random Access Memories}
\acrodefplural{RRAM}[R-RAMs]{Resistive Random Access Memories}
\acrodefplural{STT-MRAM}[STT-MRAMs]{Spin-Transfer Torque Magnetic Random Access Memories}
\acrodef{ADC}[ADC]{Analog to Digital Converter}
\acrodef{ADEX}[AdExp-I\&F]{Adaptive-Exponential Integrate and Fire}
\acrodef{AER}[AER]{Address-Event Representation}
\acrodef{AEX}[AEX]{AER EXtension board}
\acrodef{AE}[AE]{Address-Event}
\acrodef{AFM}[AFM]{Atomic Force Microscope}
\acrodef{AGC}[AGC]{Automatic Gain Control}
\acrodef{AMDA}[AMDA]{AER Motherboard with D/A converters}
\acrodef{ANN}[ANN]{Attractor Neural Network}
\acrodef{API}[API]{Application Programming Interface}
\acrodef{ARM}[ARM]{Advanced RISC Machine}
\acrodef{ASIC}[ASIC]{Application Specific Integrated Circuit}
\acrodef{BCM}[BMC]{Bienenstock-Cooper-Munro}
\acrodef{BD}[BD]{Bundled Data}
\acrodef{BEOL}[BEOL]{Back-end of Line}
\acrodef{BG}[BG]{Bias Generator}
\acrodef{BMI}[BMI]{Brain-Machince Interface}
\acrodef{CAD}[CAD]{Computer Aided Design}
\acrodef{CAM}[CAM]{Content Addressable Memory}
\acrodef{CAVIAR}[CAVIAR]{Convolution AER Vision Architecture for Real-Time}
\acrodef{CFC}[CFC]{Current to Frequency Converter}
\acrodef{CCN}[CCN]{Cooperative and Competitive Network}
\acrodef{CDR}[CDR]{Clock-Data Recovery}
\acrodef{CFC}[CFC]{Current to Frequency Converter}
\acrodef{CHP}[CHP]{Communicating Hardware Processes}
\acrodef{CNN}[CCN]{Convolutional Neural Network}
\acrodef{CMIM}[CMIM]{Metal-insulator-metal Capacitor}
\acrodef{CML}[CML]{Current Mode Logic}
\acrodef{CMOL}[CMOL]{``Hybrid CMOS nanoelectronic circuits''}
\acrodef{CMOS}[CMOS]{Complementary Metal-Oxide-Semiconductor}
\acrodef{CNN}[CCN]{Convolutional Neural Network}
\acrodef{COTS}[COTS]{Commercial Off-The-Shelf}
\acrodef{CPG}[CPG]{Central Pattern Generator}
\acrodef{CPLD}[CPLD]{Complex Programmable Logic Device}
\acrodef{CPU}[CPU]{Central Processing Unit}
\acrodef{CV}[CV]{Coefficient of Variation}
\acrodef{DAC}[DAC]{Digital to Analog Converter}
\acrodef{DAS}[DAS]{Dynamic Auditory Sensor}
\acrodef{DAVIS}[DAVIS]{Dynamic and Active Pixel Vision Sensor}
\acrodef{DBN}[DBN]{Deep Belief Network}
\acrodef{DFA}[DFA]{Deterministic Finite Automaton}
\acrodef{DI}[DI]{Delay Insensitive}
\acrodef{DMA}[DMA]{Direct Memory Access}
\acrodef{DNF}[DNF]{Dynamic Neural Field}
\acrodef{DNN}[DNN]{Deep Neural Network}
\acrodef{DOF}[DOF]{Degrees of Freedom}
\acrodef{DPE}[DPE]{Dynamic Parameter Estimation}
\acrodef{DPI}[DPI]{Differential Pair Integrator}
\acrodef{DR-RZ}[DR-RZ]{Dual-Rail Return-to-Zero}
\acrodef{DRAM}[DRAM]{Dynamic Random Access Memory}
\acrodef{DR}[DR]{Dual Rail}
\acrodef{DSP}[DSP]{Digital Signal Processor}
\acrodef{DVS}[DVS]{Dynamic Vision Sensor}
\acrodef{EBL}[EBL]{Electron Beam Lithography}
\acrodef{EDVAC}[EDVAC]{Electronic Discrete Variable Automatic Computer}
\acrodef{EIN}[EIN]{Excitatory-Inhibitory Network}
\acrodef{EM}[EM]{Expectation Maximization}
\acrodef{EPSC}[EPSC]{Excitatory Post-Synaptic Current}
\acrodef{EPSP}[EPSP]{Excitatory Post-Synaptic Potential}
\acrodef{FDSOI}[FD-SOI]{Fully-Depleted Silicon on Insulator}
\acrodef{FET}[FET]{Field-Effect Transistor}
\acrodef{FFT}[FFT]{Fast Fourier Transform}
\acrodef{FI}[F-I]{Frequency-Current}
\acrodef{FPGA}[FPGA]{Field Programmable Gate Array}
\acrodef{FSA}[FSA]{Finite State Automaton}
\acrodef{FSM}[FSM]{Finite State Machine}
\acrodef{GOPS}[GOPS]{Giga-Operations per Second}
\acrodef{GPU}[GPU]{Graphical Processing Unit}
\acrodef{GUI}[GUI]{Graphical User Interface}
\acrodef{HAL}[HAL]{Hardware Abstraction Layer}
\acrodef{HH}[H\&H]{Hodgkin \& Huxley}
\acrodef{HMM}[HMM]{Hidden Markov Model}
\acrodef{HRS}[HRS]{High-Resistive State}
\acrodef{HR}[HR]{Human Readable}
\acrodef{HSE}[HSE]{Handshaking Expansion}
\acrodef{HW}[HW]{Hardware}
\acrodef{ICT}[ICT]{Information and Communication Technology}
\acrodef{IC}[IC]{Integrated Circuit}
\acrodef{IF2DWTA}[IF2DWTA]{Integrate \& Fire 2--Dimensional WTA}
\acrodef{IFSLWTA}[IFSLWTA]{Integrate \& Fire Stop Learning WTA}
\acrodef{IF}[I\&F]{Integrate-and-Fire}
\acrodef{IMU}[IMU]{Inertial Measurement Unit}
\acrodef{INCF}[INCF]{International Neuroinformatics Coordinating Facility}
\acrodef{INI}[INI]{Institute of Neuroinformatics}
\acrodef{IO}[I/O]{Input/Output}
\acrodef{IoT}[IoT]{Internet of Things}
\acrodef{IPSC}[IPSC]{Inhibitory Post-Synaptic Current}
\acrodef{IPSP}[IPSP]{Inhibitory Post-Synaptic Potential}
\acrodef{IP}[IP]{Intellectual Property}
\acrodef{ISI}[ISI]{Inter-Spike Interval}
\acrodef{IoT}[IoT]{Internet of Things}
\acrodef{JFLAP}[JFLAP]{Java - Formal Languages and Automata Package}
\acrodef{LEDR}[LEDR]{Level-Encoded Dual-Rail}
\acrodef{LFP}[LFP]{Local Field Potential}
\acrodef{LLC}[LLC]{Low Leakage Cell}
\acrodef{LNA}[LNA]{Low-Noise Amplifier}
\acrodef{LPF}[LPF]{Low-Pass Filter}
\acrodef{LRS}[LRS]{Low-Resistive State}
\acrodef{LSM}[LSM]{Liquid State Machine}
\acrodef{LTD}[LTD]{Long Term Depression}
\acrodef{LTI}[LTI]{Linear Time-Invariant}
\acrodef{LTP}[LTP]{Long Term Potentiation}
\acrodef{LTU}[LTU]{Linear Threshold Unit}
\acrodef{LUT}[LUT]{Look-Up Table}
\acrodef{LVDS}[LVDS]{Low Voltage Differential Signaling}
\acrodef{MCMC}[MCMC]{Markov-Chain Monte Carlo}
\acrodef{MEMS}[MEMS]{Micro Electro Mechanical System}
\acrodef{MIM}[MIM]{Metal Insulator Metal}
\acrodef{MOSCAP}[MOSCAP]{Metal Oxide Semiconductor Capacitor}
\acrodef{MOSFET}[MOSFET]{Metal Oxide Semiconductor Field-Effect Transistor}
\acrodef{MOS}[MOS]{Metal Oxide Semiconductor}
\acrodef{MRI}[MRI]{Magnetic Resonance Imaging}
\acrodef{NDFSM}[NDFSM]{Non-deterministic Finite State Machine} 
\acrodef{ND}[ND]{Noise-Driven}
\acrodef{NEF}[NEF]{Neural Engineering Framework}
\acrodef{NHML}[NHML]{Neuromorphic Hardware Mark-up Language}
\acrodef{NIL}[NIL]{Nano-Imprint Lithography}
\acrodef{NMDA}[NMDA]{N-Methyl-D-Aspartate}
\acrodef{NME}[NE]{Neuromorphic Engineering}
\acrodef{NRZ}[NRZ]{Non-Return-to-Zero}
\acrodef{NSM}[NSM]{Neural State Machine}
\acrodef{OTA}[OTA]{Operational Transconductance Amplifier}
\acrodef{PCB}[PCB]{Printed Circuit Board}
\acrodef{PCHB}[PCHB]{Pre-Charge Half-Buffer}
\acrodef{PE}[PE]{Phase Encoding}
\acrodef{PCM}[PCM]{Phase Change Memory}
\acrodef{PFM}[PFM]{Pulse Frequency Modulation}
\acrodef{PR}[PR]{Production Rule}
\acrodef{PSC}[PSC]{Post-Synaptic Current}
\acrodef{PSTH}[PSTH]{Peri-Stimulus Time Histogram}
\acrodef{QDI}[QDI]{Quasi Delay Insensitive}
\acrodef{RAM}[RAM]{Random Access Memory}
\acrodef{RELU}[ReLu]{Rectified Linear Unit}
\acrodef{RLS}[RLS]{Recursive Least-Squares}
\acrodef{RMSE}[RMSE]{Root Mean Squared-Error}
\acrodef{RMS}[RMS]{Root Mean Squared}
\acrodef{RNN}[RNN]{Recurrent Neural Network}
\acrodef{ROLLS}[ROLLS]{Reconfigurable On-Line Learning Spiking}
\acrodef{RRAM}[R-RAM]{Resistive Random Access Memory}
\acrodef{SAC}[SAC]{Selective Attention Chip}
\acrodef{SCX}[SCX]{Silicon CorteX}
\acrodef{SD}[SD]{Signal-Driven}
\acrodef{SEM}[SEM]{Spike-based Expectation Maximization}
\acrodef{SLAM}[SLAM]{Simultaneous Localization and Mapping}
\acrodef{SOC}[SOC]{System-On-Chip}
\acrodef{SOI}[SOI]{Silicon on Insulator}
\acrodef{SRAM}[SRAM]{Static Random Access Memory}
\acrodef{STDP}[STDP]{Spike-Timing Dependent Plasticity}
\acrodef{STD}[STD]{Short-Term Depression}
\acrodef{STP}[STP]{Short-Term Plasticity}
\acrodef{STT-MRAM}[STT-MRAM]{Spin-Transfer Torque Magnetic Random Access Memory}
\acrodef{STT}[STT]{Spin-Transfer Torque}
\acrodef{SW}[SW]{Software}
\acrodef{TCAM}[TCAM]{Ternary Content-Addressable Memory}
\acrodef{TFT}[TFT]{Thin Film Transistor}
\acrodef{USB}[USB]{Universal Serial Bus}
\acrodef{VHDL}[VHDL]{VHSIC Hardware Description Language}
\acrodef{VLSI}[VLSI]{Very Large Scale Integration}
\acrodef{VOR}[VOR]{Vestibulo-Ocular Reflex}
\acrodef{WTA}[WTA]{Winner-Take-All}
\acrodef{XML}[XML]{eXtensible Mark-up Language}
\acrodef{divmod3}[DIVMOD3]{divisibility of a number by 3}
\acrodef{hWTA}[hWTA]{Hard Winner-Take-All}
\acrodef{sWTA}[sWTA]{soft Winner-Take-All}